\documentclass[twocolumn,showpacs,prl]{revtex4}
\usepackage{graphicx}
\usepackage{dcolumn}
\usepackage{amsmath}
\usepackage{amssymb}
\usepackage{times}

\makeatletter
\def\btt#1{\texttt{\@backslashchar#1}}
\DeclareRobustCommand\bblash{\btt{\@backslashchar}} \makeatother

\begin{document}

\title[Short Title]{
Conductance Spectroscopy of Spin-triplet Superconductors
}
\author{Yasuhiro Asano$^1$, Yukio Tanaka$^{2}$, Alexander A. Golubov$^3$, and
Satoshi Kashiwaya$^4$%
}
\affiliation{$^1$Department of Applied Physics, Hokkaido University, Sapporo 060-8628,
Japan\\
$^2$CREST-JST and Department of Applied Physics, Nagoya University, Nagoya 464-8603, Japan\\
$^3$Faculty of Science and Technology, University of Twente, 7500 AE,
Enschede, The Netherlands \\
$^4$National Institute of Advanced Industrial Science and Technology,
Tsukuba 305-8568, Japan}
\date{\today}

\begin{abstract}
We propose a novel experiment to identify the symmetry of superconductivity
on the basis of theoretical results for differential conductance of a normal metal
connected to a superconductor.
The proximity effect from the superconductor modifies the conductance of
the remote current depending remarkably on the pairing symmetry: spin-singlet or spin-triplet.
The clear-cut difference in the conductance is explained by symmetry of Cooper pairs
in a normal metal with respect to frequency. In the spin-triplet case, 
the anomalous transport is realized due to an odd-frequency symmetry 
of Cooper pairs.
\end{abstract}

\pacs{74.50.+r, 74.25.Fy,74.70.Tx}
\maketitle
The clear distinction between spin-singlet and spin-triplet superconductors is
currently a challenging issue in condensed matter physics.
Several experimental methods can be used for this purpose: the nuclear magnetic resonance,
the muon spin rotation, the critical magnetic field beyond the Pauli limit,
the Josephson $\pi$-junctions, the thermal conductivity, and the detection of
multiple-phases by the spin susceptibility and the specific heat.
For instance, the unchanged Knight-shift across the critical temperature $T_c$
suggests the spin-triplet superconductivity.
This result, however, is not a sufficient condition for the spin-triplet superconductivity
because the spin-singlet superconductivity with the strong spin-orbit coupling may
also explain the unchanged Knight-shift.
In addition, the complicated procedure of the data analysis may make the conclusion unclear.
Other experiments also involve such unclear factors.
At present, only a series of different experiments can lead to the conclusion about
the spin-triplet superconductivity as in the case of Sr$_2$RuO$_4$~\cite{mackenzie}.
%
The experimental methods listed above involve an applied magnetic field. Such
experiments are powerless to analyze the ferromagnetic superconductors such as
UGe$_2$ and URhGe~\cite{saxena} because the magnetic moment of
a superconductor spoils an experimental signal.
However these compounds are undoubtedly promising candidates for the spin-triplet superconductivity.
To find a way out of the stalemate, one has to use intrinsic phenomena related to the spin-triplet
superconductivity. We address this issue in the present paper.

\begin{figure}[tbh]
\begin{center}
\includegraphics[width=7.5cm]{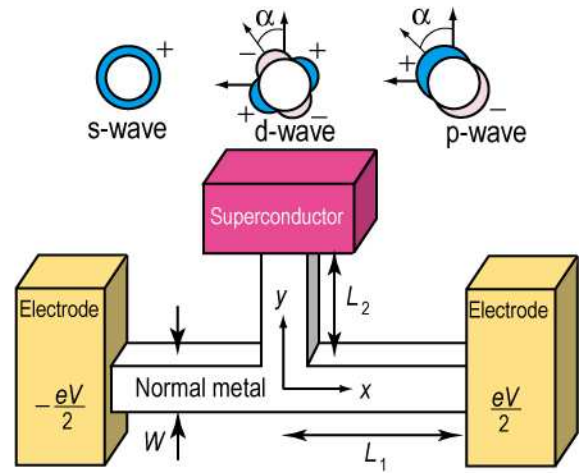}
\end{center}
\caption{ (Color online) Schematic figures of a T-shaped proximity structure and
pair potentials on the Fermi surface.}
\label{fig1}
\end{figure}

In normal-metal/superconductor (NS) junctions, Cooper pairs penetrate from a superconductor
into a normal metal. This phenomenon is called the proximity effect and is very sensitive to
the pairing symmetry of a superconductor~\cite{ya01-2,yt03-1,ya06-1,yt04}.
In junctions with spin-triplet superconductors, Cooper pairs penetrating into a normal metal have 
the odd-frequency symmetry~\cite{tanaka07}.
Although the odd-frequency superconductivity itself has never been confirmed in any material,
the proximity effect involving odd-frequency pairs is currently a hot topic~\cite{bergeret,keizer,tanaka07,ya07}.
The existence of odd-frequency pairs causes the drastic enhancement of
the quasiparticle density of states (DOS) at the Fermi energy
in a normal metal~\cite{tanaka05r,ya07}.
We will show in this paper that the enhancement of the DOS can be detected as a zero-bias anomaly
in the differential conductance of a proximity structure as shown in Fig.~\ref{fig1}.
In the last decade, several authors discussed interesting feature of
the proximity effect on the remote electric current in similar junctions
of $s$-wave superconductor~\cite{nazarov2,volkov}.
We will also show that the conductance spectra always have a dip structure around the
zero-bias for all spin-singlet superconductor junctions.
On the basis of the robustness of the phenomena, we will conclude that the
conductance spectroscopy may serve as a useful tool to test the spin-triplet superconductivity.
This discussion connects the physics of mesoscopic transport and that of unconventional
superconductivity.

Let us consider the T-shaped junction as shown in Fig.~\ref{fig1}.
The bias-voltage $eV$ is applied to the horizontal normal metal which is
connected with two electrodes at $x=\pm L_1$.
The normal metal has the third branch which is terminated by a superconductor at $y=L_2$.
%
To calculate the conductance of a normal metal, we solve the quasiclassical Usadel
equation~\cite{usadel} in the Keldysh formalism,
\begin{align}
\hbar D \nabla &\left\{ \check{G}(\boldsymbol{r}) \nabla \check{G}(\boldsymbol{r}) \right\}
+ i \left[ \check{H}, \check{G}(\boldsymbol{r})\right]_{-} =0,\label{usa1}\\
\check{G}(\boldsymbol{r})=&
\left(\begin{array}{cc}
\hat{g}^R(\boldsymbol{r}) & \hat{g}^K(\boldsymbol{r}) \\
\hat{0} & \hat{g}^A(\boldsymbol{r})
\end{array}\right),
\check{H}= \left(\begin{array}{cc}
\epsilon \hat{\tau}_3 & \hat{0} \\
\hat{0} & \epsilon \hat{\tau}_3
\end{array}\right),
\end{align}
where $D$ is the diffusion constant of a normal metal, $\epsilon$ is the energy of a
quasiparticle measured from the Fermi level, and $\hat{\tau}_i$ for $i=1-3$ are the Pauli matrices.
The symbols $\hat{\cdots}$ and $\check{\cdots}$ indicate $2\times 2$ and $4\times 4$ matrices,
respectively. In the following, we solve the Usadel equation in two dimensions.
The results are valid also for three-dimensional junctions shown in Fig.~\ref{fig1}.
We assume that a spin-triplet Cooper pair
consists of two electrons with the opposite spin directions.
This assumption does not break the generality of the following discussion.
The Keldysh Green function can be decomposed by $
\hat{g}^K=\hat{g}^R \hat{h} - \hat{h} \hat{g}^A$ with $\hat{h}= f_L + f_T \hat{\tau}_3$,
where $f_L$ and $f_T$ are the distribution function of a quasiparticle.
From the Keldysh part of Eq.~(\ref{usa1}), we derive the modified diffusion
equation which describes the kinetics of a quasiparticle in a normal metal~\cite{belzig,volkov2},
\begin{align}
\nabla \left( D_T \nabla f_T \right) = 0, \label{md}
\end{align}
with $D_T= \text{Tr}\left( 1- \hat{g}^R \hat{\tau}_3 \hat{g}^A \hat{\tau}_3 \right)/4$.
The electric current defined by $I= eN_0 D \int_{-\infty}^\infty \!\! d\epsilon \; J_T$
can be calculated from the integration of
Eq.~(\ref{md}) along $\boldsymbol{r}=(0,0)$ to $\boldsymbol{r}=(L_1,0)$,
\begin{align}
J_T = \frac{F_R}{ L_1^{-1}\int_0^{L_1} \!\! dx\; D_T^{-1}},
\end{align}
where we apply the boundary conditions as
\begin{align}
f_T(x=L_1)=F_R=\frac{1}{2}
\left\{
\tanh \left( \frac{\epsilon_+}{2T} \right) -
\tanh \left( \frac{\epsilon_-}{2T} \right) \right\}
\end{align}
with $\epsilon_\pm=\epsilon\pm eV/2$ and $T$ being a temperature.
At $x=0$, we also apply $f_T=0$.
At the cross-point $\boldsymbol{r}=(0,0)$, the current conservation low implies
$ \left.\sum_i \boldsymbol{n}_i \check{G} \nabla_i \check{G}\right|_{\boldsymbol{r}=0} =0$,
 where $\boldsymbol{n}_i$
is the unit vector points to outside of the cross-point~\cite{zaitsev}.
The retarded part of the Usadel equation is expressed by the usual $\theta$-parameterization
\begin{align}
\hbar D \nabla^2 \theta(\boldsymbol{r}) + 2i \epsilon \sin \theta(\boldsymbol{r})=0.\label{usa2}
\end{align}
We find the relation $D_T=\cosh^2\left[ \text{Im} \theta(\boldsymbol{r})\right]$.
The Usadel equation is supplemented by the boundary condition at $\boldsymbol{r}=(0,L_2)$
which depends on the pairing symmetry of a superconductor~\cite{nazarov,yt03-1,yt04},
\begin{align}
&\left.\frac{\partial \theta(0,y)}{\partial y}\right|_{\boldsymbol{r}=(0,L_2)}
= \frac{\rho_N}{W} \frac{\langle F \rangle}{R_B T_B},\label{naz}\\
\langle F \rangle =& \int_{-\pi/2}^{\pi/2} \!\!\!\!\! d\gamma
\frac{ T_N \cos\gamma  (f_s \cos\theta_0 - g_s \sin\theta_0)}
{(2- T_N)\xi + T_N ( g_s \cos\theta_0 + f_s \sin\theta_0)}, \label{deff}
\end{align}
where $T_N={ \cos^2\gamma}/({z_0^2 + \cos^2\gamma})$,
$\gamma$ is the incident angle of
a quasiparticle measured from the $y$ axis,
$\theta_0=\theta(0, L_2)$, and $\rho_N$ is the resistivity of a normal metal.
The transmission probability and the resistance of the NS interface are given by
$T_B = \int_0^{\pi/2}\!\!\! d\gamma \cos\gamma\; T_N$ and
 $R_B=[(2e^2/h)(k_FW)T_B/\pi]^{-1}$, respectively.
The Green function in a superconductor depends on $\gamma$ and the orientation angle
$\alpha$ in Fig.~\ref{fig1} as
$g_\pm = \epsilon / \sqrt{ \epsilon^2 - \Delta_\pm^2}$
and $f_\pm = \Delta_\pm / \sqrt{\Delta_\pm^2 - \epsilon^2}$, where
$\Delta_\pm=\Delta_0\Psi(\gamma_\pm)$ with $\Delta_0$ being the amplitude of the
pair potential at $T=0$,  $\gamma_+=\gamma-\alpha$ and $\gamma_-=\pi -\gamma -\alpha$.
The form factor $\Psi(\gamma)$ characterizes the pairing symmetry as
$\Psi(\gamma)=1$, $\cos \gamma$, and $\cos 2\gamma$ for $s$-, $p$-, and $d$-wave
symmetries, respectively. In Eq.~(\ref{naz}), $g_s=(g_++g_-)$,
$\xi=1+g_+g_-+f_+f_-$, and $f_s=(f_++f_-)$ for the spin-singlet pairing symmetry
 and $f_s= i(f_+g_- - f_-g_+)$ for the spin-triplet one~\cite{yt03-1,yt04}.
At $x=\pm L_1$, we impose $\theta(\pm L_1,0)=0$.
The differential conductance at zero temperature results in
\begin{align}
R_N\left.\frac{dI}{dV}\right|_{eV} =\!\!\! \left.
\left[ \frac{1}{L_1} \int_0^{L_1}\!\! \frac{dx}{\cosh^2\left[ \text{Im} \theta(x,0)\right]}
\right]^{-1}\right|_{\epsilon=eV}, \label{didv}
\end{align}
where $R_N=2L_1\rho_N/W$ is the normal state resistance of the junction.
In what follows, we fix the Thouless energy of a half horizontal wire
$E_{th}\equiv \hbar D/L_1^2$ at 0.04$\Delta_0$.

\begin{figure}[tbh]
\begin{center}
\includegraphics[width=8.0cm]{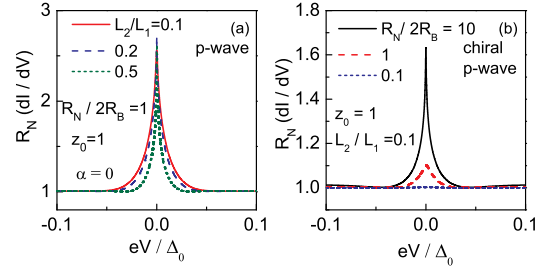}
\end{center}
\caption{ (Color online) The differential conductance in the $p$-wave
symmetry in (a) and the chiral $p$-wave symmetry in (b).
}
\label{fig2}
\end{figure}

First we discuss the differential conductance of the spin-triplet $p$-wave junctions
in Fig.~\ref{fig2}(a), where $R_N/2R_B=1$, $z_0=1$, and $\alpha=0$.
The conductance has strong zero-bias peak.
The width of the zero-bias conductance peak (ZBCP) decreases with increasing $L_2$ because
the energy $\hbar D/(L_1+L_2)^2$ characterizes the peak width.
Thus one should fabricate $L_2$ as short as possible to observe the ZBCP in experiments.
In what follows, we fix $L_2/L_1=0.1$.
On the other hand, the height of the ZBCP is independent of $L_2$ as shown
in Fig.~\ref{fig2}(a) and follows from the analytic expression of the zero-bias conductance,
\begin{align}
R_N\left.\frac{dI}{dV}\right|_{eV=0} = \frac{ \frac{R_N \cos\alpha}{2R_BT_B} }{
\tanh\left(  \frac{R_N \cos\alpha}{2R_BT_B} \right)}. \label{zbcp}
\end{align}
The amplitude of the ZBCP decreases with the increase of the orientation angle $\alpha$
and vanishes at $\alpha=\pi/2$. This is because the proximity effect is absent
in a normal metal at $\alpha=\pi/2$~\cite{ya01-2,yt04}.
In the $p$-wave symmetry case, the ZBCP can be observed at temperatures below $E_{th}$ 
for almost all orientation angles.
In Fig.~\ref{fig2}(b), we discuss the conductance in the chiral $p$-wave symmetry
to test realistic junctions involving Sr$_2$RuO$_4$~\cite{mackenzie},
where the form factor is given by $\Psi(\gamma) = \cos\gamma + i \sin\gamma$.
The boundary condition in Eq.~(\ref{deff}) can be used with
$g_s=\{ 2g +i (f_{1+}f_{2-}-f_{2+}f_{1-}) \}$,
$f_s=- \{ ig(f_{1+}-f_{1-}) + f_{2+}+f_{2-} \}$,
$\xi=1+g^2+f_{1+}f_{1-}+f_{2+}f_{2-}$,
$g=\epsilon/\sqrt{\epsilon^2-\Delta_0^2}$, and
$f_{1(2),\pm}=\text{Re(Im)} \Delta_\pm / \sqrt{\Delta_0^2-\epsilon^2}$~\cite{tanaka05r}.
In the chiral $p$-wave junctions, the peak width is characterized also by $E_{th}$ but
the zero-bias conductance is approximately
given by Eq.~(\ref{zbcp}) with $\cos\alpha/T_B \to 1$.
In the limit of weak proximity effect such as $R_N/2R_B=0.1$, the ZBCP
becomes small. In other cases, the proximity effect leads to the clear ZBCP
as in Fig.~\ref{fig2}(a). The conductance spectra in Fig.~\ref{fig2}(a) and
(b) show qualitatively similar behavior.
In spin-triplet junctions, the boundary condition in Eq.~(\ref{deff}) and the Usadel
equation in Eq.~(\ref{usa2}) at $\epsilon=0$ yield
the pure imaginary value of $\theta$ everywhere in a normal metal.
Then the zero-bias anomaly in the conductance follows mathematically from this fact 
and Eq.~(\ref{didv}) and therefore is the robust feature of the spin-triplet
superconductors.

\begin{figure}[tbh]
\begin{center}
\includegraphics[width=8.0cm]{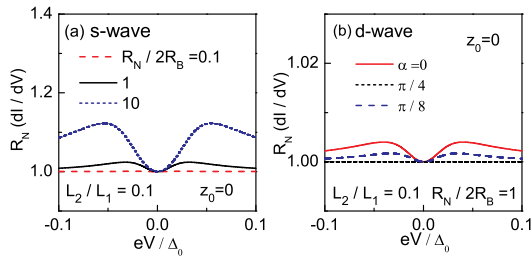}
\end{center}
\caption{ (Color online) The differential conductance in the spin-singlet
superconductor junctions for $s$-wave symmetry in (a) and $d$-wave symmetry in (b).
}
\label{fig3}
\end{figure}
Next, let us summarize the differential conductance of the spin-singlet superconducting
junctions in Fig.~\ref{fig3}. In Fig.~\ref{fig3}(a), the results for $s$-wave junctions
are plotted for several choices of $R_N/2R_B$ at $z_0=0$.
In contrast to the spin-triplet cases in Figs.~\ref{fig2}, the conductance has the dip structure.
In the spin-singlet junctions, the proximity effect has two contributions which influence
the conductance in an opposite way.
The induced superconductivity in a normal metal tend to assist the electron transport.
On the other hand, the existence of Cooper pairs decreases the DOS in a normal
metal (the so-called minigap is formed), and this leads to the suppression of the conductance.
These two effects exactly cancel each other at $eV=0$~\cite{volkov}.
The positive contribution to the conductance due the proximity effect 
decays in power low of $eV$, whereas the negative contribution decays
exponentially~\cite{volkov}. Thus the proximity effect slightly enhances
the conductance around $eV \sim E_{th}$ and the conductance spectra show the
dip structure as shown in Fig.~\ref{fig3}~\cite{nazarov2,volkov}.
The degree of the enhancement depends on the strength of the proximity effect
($R_N/2R_B$) and is typically of the order of percent as shown in (a).
The characteristic behavior of the conductance is insensitive to $z_0$.
In Fig.~\ref{fig3} (b), we show the conductance in the $d$-wave symmetry for several choices of the orientation
angle $\alpha$, where $z_0=0$ and $R_N/2R_B=1$. At $\alpha=0$, the conductance shows the dip
structure near the zero-bias as well as those in the $s$-wave junctions.
The dip structure gradually disappears with the increase of $\alpha$.
The conductance spectra become completely flat at $\alpha=\pi/4$ because
the proximity effect is absent in a normal metal~\cite{ya01-2,yt03-1}.
In contrast to the spin-triplet junctions,
Eqs.~(\ref{usa2}) and (\ref{naz}) at $\epsilon=0$ always yield
real value of $\theta$ for all spin-singlet junctions.
This fact mathematically explains the cancellation of the two contributions of the
proximity effect at $eV=0$ because Eq.~(\ref{didv}) results in $dI/dV=R_N^{-1}$
for real $\theta$.
Thus the dip structure around the zero-bias in the conductance spectra is
the robust feature of the spin-singlet superconductor junctions.
This conclusion is valid only for the T-shaped junction in which a superconductor is away
from the current path. In usual quasi one-dimensional NS junctions,
the proximity effect causes the ZBCP in the $s$-wave symmetry~\cite{kastalsky}.

Here we explain the reasons of the zero-bias anomaly in the spin-triplet junctions.
Since electrons obeys the Fermi statistics, the pairing function of a Cooper pair
satisfies the relation
\begin{align}
f_{\sigma,\sigma'}(\boldsymbol{k},\epsilon)
= - f_{\sigma',\sigma}(-\boldsymbol{k},-\epsilon),\label{pauli}
\end{align}
where $\sigma$ and $\sigma'$ are the spin of the two electrons,
$\boldsymbol{k}$ dependence of pairing function
is the characterizes the symmetry of the orbital part.
According to the relation, the
Cooper pair in a superconductor is classified into the spin-singlet even-parity
 and the spin-triplet odd-parity symmetry classes.
The interchange of spin (i.e., $\sigma \leftrightarrow \sigma'$)
and $\boldsymbol{k} \to -\boldsymbol{k}$ give rise to the
negative sign on the right hand side of Eq.~(\ref{pauli}) in the former
and in the latter, respectively.
In a normal metal, only $s$-wave pairs are allowed
irrespective of an original pairing symmetry in a superconductor
because of the diffusive impurity scattering.
In the $p$-wave junctions, the spin-triplet $s$-wave Cooper pairs
penetrate into a normal metal.
To satisfy Eq.~(\ref{pauli}), such Cooper pairs
acquire the odd-frequency symmetry property (i.e.,
$f_{\sigma,\sigma'}(\boldsymbol{k},\epsilon)
= - f_{\sigma,\sigma'}(\boldsymbol{k},-\epsilon)$).
The most important feature of the odd-frequency pairs is the enhancement
of the quasiparticle DOS at $\epsilon=0$~\cite{tanaka05r,ya07}.
This feature is in contrast to the usual proximity effect in the spin-singlet junctions.
Thus the proximity effect always increases the conductance in the spin-triplet junctions.
In addition to this, the DOS at $\epsilon=0$ becomes large because of
the midgap Andreev resonant state (MARS)~\cite{hu,tanaka95}.
The large DOS at the Fermi energy is interpreted as the penetration
of the MARS from a superconductor into a normal metal~\cite{yt04,ya06-1}.
Thus the ZBCP in Fig.~\ref{fig2} reflects the peak structure of the DOS in a normal metal.
The effects of MARS in the chiral $p$-wave symmetry are weaker than those in
the $p$-wave symmetry because only quasiparticles with $\gamma=0$ contribute to the MARS.
Thus the zero-bias conductance in the chiral $p$-wave junction is smaller than that in the
$p$-wave as shown in Fig.~\ref{fig2}.
The odd-frequency symmetry compensates the symmetry change of the orbital part
from the odd-parity symmetry in a superconductor to the $s$-wave symmetry in a normal metal.
Therefore we conclude that the ZBCP is expected in the T-shaped junctions of
all spin-triplet superconductors independent of detailed structures in
$\boldsymbol{d}$-vector.

 Finally we propose a new experiment to discriminate the symmetry of a superconductor.
The proximity effect on the remote current causes the clear-cut difference
between the conductance spectra of the spin-triplet junctions and those
in the spin-singlet ones as shown in Figs.~\ref{fig2} and \ref{fig3}.
Therefore the T-shaped junction can serve as a superconducting symmetry detector.
In addition, the present proposal has several advantages compared to the previous proposals~\cite{ya06-1,yt04}
with respect to resolving the low energy transport .
First, to observe the characteristic conductance spectra at $|eV| \lesssim E_{th}$,
it is necessary to suppress the influence of undesired
scattering due to defects and/or localized states at the NS
interface because the tunneling conductance is extremely sensitive to the interface quality.
The tunneling current through such states easily washes out the
expected conductance signals. In fact, the bad interface quality
damages the subgap tunneling spectra of Sr$_2$RuO$_4$.
The conductance of the T-shaped geometry, however, is rather insensitive to the interface
quality because the current does not flow through the NS interface.
Second, within the present technologies it is difficult to realize small and highly transparent NS
junctions for observing the Josephson current.
This is because the unconventional superconductors are usually synthesized as bulk
materials and are not suitable for microfabrication.
The T-shaped junctions, however, requires the
microfabrication only on the normal metal (not on a superconductor).
Thus the T-shaped junctions are accessible within the present technique.
Finally the proposed experiment can test the ferromagnetic superconductors because
the measurement of the conductance spectra does not require external magnetic field.
For these reasons, we conclude that the T-shaped junctions would be a powerful
tool to test the symmetry of superconductors.

In conclusion, we have studied the conductance spectra of T-shaped superconductor
junctions. The proximity effect on the remote current modifies the low energy
transport depending remarkably on the symmetry of superconductors.
In the case of the spin-triplet superconductors, the conductance shows the zero-bias anomaly.
The odd-frequency Cooper pairs in a normal metal cause the anomaly and the midgap
Andreev resonant states support the robustness of this drastic effect.
In contrast to the spin-triplet case, the conductance spectra in the spin-singlet
junctions always show the dip structure around the zero-bias.
On the basis of the calculation results,
we have proposed a new experimental method to detect spin-triplet superconductivity
and have discussed the advantages of the method.

This work was partially supported by the
Dutch FOM, the NanoNed program under grant TCS7029 and
Grant-in-Aid for Scientific
Research from The Ministry of Education,
Culture, Sports, Science and Technology of Japan
(Grant No. 18043001, 17071007, 19540352 and 17340106).


\begin{thebibliography}{99}

\bibitem{mackenzie} A.~P.~Mackenzie and Y.~Maeno, Rev. Mod. Phys. \textbf{75}, 657 (2003).


\bibitem{saxena} S.~S.~Saxena, et. al. Nature \textbf{406}, 587 (2000);
D.~Aoki, et. al., Nature \textbf{413}, 613 (2001).

\bibitem{ya01-2} Y.~Asano, Phys. Rev. B \textbf{64}, 014511 (2001);
J. Phys. Soc. Jpn. \textbf{71}, 905 (2002).

\bibitem{yt03-1} Y.~Tanaka, Yu.~V.~Nazarov, and S.~Kashiwaya,
Phys. Rev. Lett. \textbf{90}, 167003 (2003).

\bibitem{ya06-1} Y.~Asano, Y.~Tanaka, and S.~Kashiwaya, Phys. Rev. Lett. \textbf{96}, 097007 (2006).
\bibitem{yt04} Y.~Tanaka and S.~Kashiwaya, Phys. Rev. B \textbf{70}, 012507 (2004);
Y.~Tanaka, S.~Kashiwaya, and T.~Yokoyama, Phys. Rev. B \textbf{71}, 094513 (2005).

\bibitem{tanaka07} Y.~Tanaka and A.~A.~Golubov, Phys. Rev. Lett. \textbf{98}, 037003 (2007).



\bibitem{bergeret} F.~S.~Bergeret, A.~F.~Volkov, and K.~B.~Efetov, Phys.
Rev. Lett. \textbf{86}, 4096 (2001); Rev. Mod. Phys. \textbf{77}, 1321
(2005).
\bibitem{keizer} R.~S.~Keizer, S.~T.~B.~Goennenwein, T.~M.~Klapwijk,
G.~Miao, G.~Xiao, A.~Gupta, Nature \textbf{439}, 825 (2006).


\bibitem{ya07} Y.~Asano, Y.~Tanaka, A.~A.~Golubov, Phys. Rev. Lett. \textbf{98}, 107002 (2007).

\bibitem{tanaka05r}Y.~Tanaka, Y.~Asano, A.~A.~Golubov, and S.~Kashiwaya,
Phys. Rev. B \textbf{72}, 140503(R) (2005).

\bibitem{nazarov2} Yu.~V.~Nazarov and T.~H.~Stoof, Phys. Rev. Lett. \textbf{76}, 823 (1996).


\bibitem{volkov} A.~F.~Volkov and H.~Takayanagi, Phys. Rev. Lett. \textbf{76}, 4026 (1996).


\bibitem{usadel} K.~Usadel, Phys. Rev. Lett. \textbf{25}, 507 (1970).

\bibitem{belzig} W.~Belzig, F.~K.~Wilhelm, C.~Bruder, and G.~Sch\"{o}n,
Superlattices and Microstructures \textbf{25}, 1251 (1999).

\bibitem{volkov2} A.~F.~Volkov, A.~V.~Zaitsev, and T.~M.~Klapwijk, Physica C \textbf{210},
21 (1993).

\bibitem{zaitsev} A.~V.~Zaitsev, Phys. Lett. A \textbf{194}, 315 (1994).

\bibitem{nazarov} Yu.~V.~Nazarov, Phys. Rev. Lett. \textbf{73}, 1420 (1994); cond-mat/9811155.


\bibitem{kastalsky} A.~Kastalsky, A.~W.~Kleinsasser, L.~H.~Greene, R.~Bhat, F.~P.~Milliken,
and J.~P.~Harbison, Phys. Rev. Lett. \textbf{67}, 3026 (1991).

\bibitem{hu} C.~R.~Hu, Phys. Rev. Lett. \textbf{72}, 1526 (1994).

\bibitem{tanaka95} Y.~Tanaka and S.~Kashiwaya, Phys. Rev. Lett. \textbf{74}, 3451 (1995).


\end{thebibliography}
\end{document}